\documentclass[noshowpacs,aps,twocolumn,pra]{revtex4-1}

\usepackage{graphicx}
\usepackage{amsmath}

\begin{document}

\author{Bogdan-Ioan Popa}
\email{bap7@ee.duke.edu}
\author{Durvesh Shinde}
\author{Adam Konneker}
\author{Steven A. Cummer}
\email{cummer@ee.duke.edu}
\affiliation{Department of Electrical and Computer Engineering, Duke University, North Carolina 27708}

\title{Active acoustic metamaterials reconfigurable in real-time}

\date{\today}

\begin{abstract} 
A major limitation of current acoustic metamaterials is that their acoustic properties are either locked into place once fabricated or only modestly tunable, tying them to the particular application for which they are designed. We present in this paper a design approach that yields active metamaterials whose physical structure is fixed, yet their local acoustic response can be changed almost arbitrarily and in real-time by configuring the digital electronics that control the metamaterial acoustic properties. We demonstrate experimentally this approach by designing a metamaterial slab configured to act as a very thin acoustic lens that manipulates differently three identical, consecutive pulses incident on the lens. Moreover, we show that the slab can be configured to implement simultaneously various roles, such as that of a lens and beam steering device. Finally, we show that the metamaterial slab is suitable for efficient second harmonic acoustic imaging devices capable to overcome the diffraction limit of linear lenses. These advantages demonstrate the versatility of this active metamaterial and highlight its broad applicability, in particular to acoustic imaging.
\end{abstract}

\maketitle

Acoustic metamaterials significantly expand our ability to manipulate sound by enabling material parameters that are not easily available in natural media, such as negative mass density, bulk modulus, and refractive index \cite{Liu_Science_2000, Fang_NatMater_2006, Lee_PRL_2010, Fok_PRB_2011, Liang_PRL_2012, Xie_PRL_2013, Garcia-Chocano_PRL_2014}, as well as highly anisotropic \cite{Torrent_NJP_2008, Pendry_NJP_2008, Popa_PRB_2009, Zigoneanu_JAP_2011} and even unidirectional acoustic properties \cite{Liang_NatMater_2010, Boechler_NatMater_2011, Popa_NatComm_2014, Fleury_Sci_2014}. Most importantly, they allow rigid materials to behave as regular fluids and are consequently suitable for robust devices for sound control. As a result, recent years have seen a remarkable development of new classes of metamaterials that are suitable for numerous applications, such as exotic sound cancellation \cite{Zhang_PRL_2011, Popa_PRL_2011, Zigoneanu_NatMater_2014} and improved acoustic imaging devices \cite{Li_NatMater_2009, Martin_APL_2010, Climente_APL_2010, Zigoneanu_PRB_2011}.

One of the drawbacks of most reported metamaterials is that their functionality is locked into place once they are fabricated. There are scenarios where it is desirable to tune their properties dynamically as needed rather than building a new device for each particular application. There are currently only a few designs that promise to address this limitation \cite{Akl_JAP_2012, Popa_PRB_2013, Popa_NatComm_2014, Fleury_NatComm_2015, Chen_APL_2014}. Thus, recent advances on active metamaterials employing electronic circuits in their structure provide a natural path towards tunable metamaterials \cite{Akl_JAP_2012, Popa_PRB_2013, Popa_NatComm_2014, Fleury_NatComm_2015}. Other approaches involve the use of variable magnetic fields to control the properties of pre-stretched membranes \cite{Chen_APL_2014}. However, these designs merely tweak a particular material parameter in a relatively narrow range. It would be advantageous to change entirely the metamaterial functionality in a dynamic manner.

We show in this paper that active acoustic metamaterials provide an ideal platform to design versatile media whose acoustic properties are easily configured dynamically. To demonstrate this concept, we fabricate a metamaterial slab and demonstrate the following two features. First, its functionality can be set in real-time without changing the physical structure of the slab. Second, it can implement acoustic properties not easily available in metamaterials designed through other methods. Acoustic imaging is a promising area of application for acoustic metamaterials \cite{Li_NatMater_2009, Martin_APL_2010, Climente_APL_2010, Zigoneanu_PRB_2011}, therefore the experimental illustration of the benefits of our design method will have these applications in mind. More specifically, to prove the first point we send a plane wave consisting of three modulated Gaussian pulses coming in quick succession, and configure the slab to act as an acoustic lens whose properties are changed rapidly so that each pulse is manipulated differently. 

To demonstrate the second point, we configure the slab to play simultaneously two different roles, namely that of a beam steering device and focusing lens. In addition, we show how very small F-numbers of less than 0.5 are possible in a very thin lens whose thickness is less than a tenth of the wavelength. Moreover, we use this opportunity to demonstrate experimentally an idea first presented in the context of electromagnetic metamaterials, namely that imaging using higher order harmonics of the incident field can beat the diffraction limit associated with linear devices.

\begin{figure}%[ht!]
  \begin{center}
    \includegraphics[width=8.5cm]{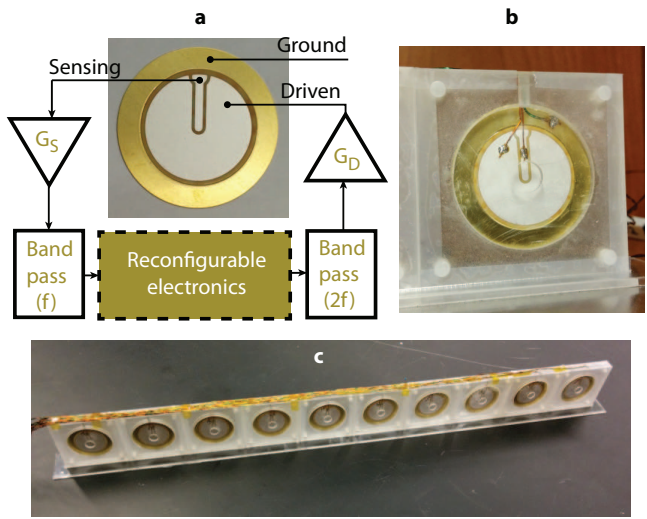}
    \caption{Reconfigurable metamaterial unit cell. (a) The unit cell consists of a piezoelectric membrane controlled by electronics. The cell acoustic response is controlled by a digital electronic circuit that can be reconfigured in real-time. (b) Photograph of fabricated unit cell. (c) Metamaterial slab consisting of ten unit cells.}
    \label{fig:unit cell}
  \end{center}
\end{figure}

The metamaterial unit cell follows the general architecture described in Refs. \onlinecite{Popa_PRB_2013, Popa_NatComm_2014}. It consists of a three terminal piezoelectric membrane produced by Murata Inc. augmented by electronics, as illustrated in Fig. \ref{fig:unit cell}a. The incident sound is sampled by the membrane's sensing terminal, which creates an electric signal proportional to the acoustic excitation. The signal is amplified in a pre-amplifier of gain $G_S$ and passes through a bandpass filter of bandwidth 20\% centered on frequency $f=1500$ Hz. It then enters a block of reconfigurable electronics that essentially determines the metamaterial functionality. Since the focus here is on second harmonic imaging applications, one of the functions of this stage is to create the second harmonic. The resulting signal passes through a second bandpass filter of bandwidth 20\% centered on $2f=3000$ Hz, and it is further amplified (gain $G_D$) and drives the main terminal of the piezoelectric membrane, thus creating the acoustic response of the cell. To increase cell repeatability and the overall flexibility of the system, the bandpass filters together with the reconfigurable electronics are implemented digitally in a digital signal processor (DSP) produced by Atmel (SAM3X8E). The cell functionality is thus set by uploading various programs into the DSP.

To better match acoustically the piezoelectric membrane to the surrounding environment, the membrane is mounted between two identical Helmholtz cavities made of Plexiglas and designed according to the membrane manufacturer recommendations. Thus, each cavity has a height 3.2 mm, a circular cross section of diameter 33 mm, and end with an opening whose diameter is 6 mm. Figure \ref{fig:unit cell}b shows a photo of the metamaterial unit cell, and Fig. \ref{fig:unit cell}c presents a metamaterial slab composed of ten identical unit cells.

The goal of this paper is to leverage the flexibility afforded by the reconfigurable electronics block and change the metamaterial functionality without changing its physical structure. The first application is to configure the slab to act as an acoustic metamaterial lens (AML) that images a distant object by focusing the plane wave incident on it into a focal spot in the lens vicinity. This has been done using passive materials and various techniques \cite{Kock_JASA_1949, Climente_APL_2010, Martin_APL_2010, Zigoneanu_PRB_2011, Wang_APL_2014}. Here we emphasize the configurability feature of this design method by changing the lens parameters, namely focal length and direction of the focused beam, in real-time as the incident wave hits the lens.

\begin{figure}%[ht!]
  \begin{center}
    \includegraphics[width=8.5cm]{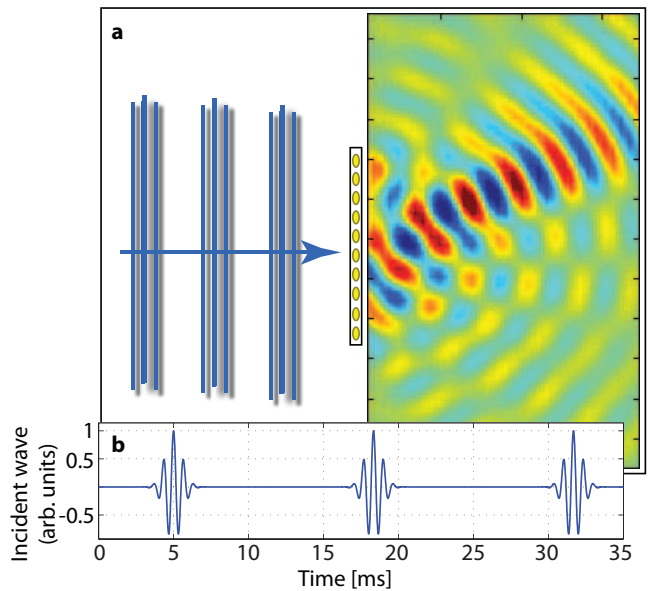}
    \caption{Experimental setup. (a) Top view sketch of the 2D acoustic waveguide showing the metamaterial slab positioned in the middle of the waveguide. The incident sound consists of a series of three identical, closely spaced pulses. The acoustic field is measured in the highlighted waveguide region behind the metamaterial in which the metamaterial response to the first pulse is shown. (b) Time domain variation of the incident sound field.}
    \label{fig:experimental setup}
  \end{center}
\end{figure}

The acoustic behavior of the AML is characterized experimentally inside a two-dimensional (2D) acoustic waveguide composed of two square parallel plastic sheets 120 cm long and separated by 5 cm. The latter dimension coincides with the metamaterial unit cell size. A top-view schematic of the waveguide showing the lens placed in the middle is presented in Fig. \ref{fig:experimental setup}a. A plane wave is obtained inside the waveguide using an array of 12 in-phase speakers placed on the left edge of the waveguide. The plane wave consists of three identical Gaussian pulses modulated by a sinusoid at 1500 Hz separated by ~13 ms. The spatial distribution of the wave is sketched in Fig. \ref{fig:experimental setup}a, and its time domain evolution at the AML position is given in Fig. \ref{fig:experimental setup}b. Our purpose is to reconfigure the lens in between the pulses so that the lens behavior is different for each pulse. More specifically, the target AML parameters for the three pulses are given by the pairs (30 cm, 30$^\circ$), (20 cm, -30$^\circ$), and (30 cm, 0$^\circ$), respectively, where the first parameter of the pair is the focal length, and the second is the angle that the focused beam makes with the horizontal.

The fields on the transmission side of the AML are measured using a microphone that scans the waveguide interior on a square grid of points separated by 2 cm in both the horizontal and vertical directions. To obtain a visually more pleasing image of the fields, the measurements are spatially interpolated to double the number of sampled points in each directions. A typical measurement is shown in Fig. \ref{fig:experimental setup}a to illustrate the scanned area relative the the lens position.

Acoustic lenses can be realized in several ways. For example, curved surfaces \cite{Belher_PO_1999} could be used to transform the incident acoustic field into a converging field. Gradient-index metamaterials could achieve the same effect by varying the delay of the acoustic wave as it propagates through various parts of a flat lens \cite{Kock_JASA_1949, Climente_APL_2010, Martin_APL_2010, Zigoneanu_PRB_2011, Wang_APL_2014}. We employ here the latter idea. However, instead of delaying the incident wave inside the material itself we engineer the delay in electronics. This approach addresses two inherent limitations of passive designs, namely lenses can be made much thinner than through other methods and the focal distance and, consequently, their F-number can be made remarkably small.

\begin{figure}%[ht!]
  \begin{center}
    \includegraphics[width=8.5cm]{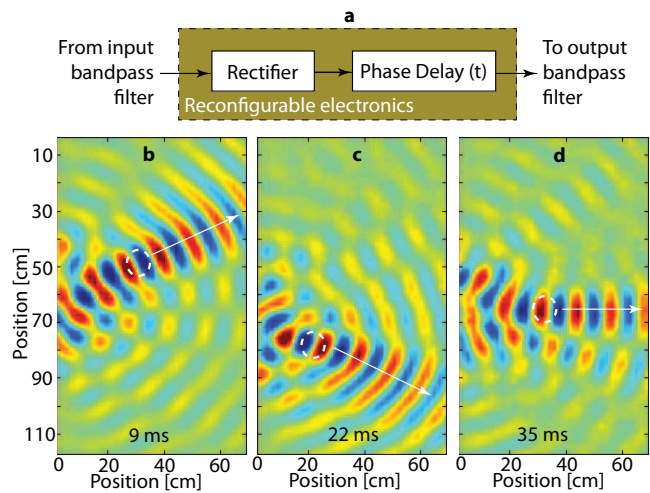}
    \caption{Lens acoustic response to three closely spaced identical pulse. (a) Functional diagram of the reconfigurable electronics block. The rectifier generates the second harmonic, and the phase delay line configures the metamaterial to behave as a lens. (b-d), Time instances of the acoustic responses to each one of the three pulses. For each pulse the metamaterial slab behaves as a lens whose parameters are reconfigured in-between pulses. The time instance, focal spots, and the direction of propagation of the beam leaving the lens are specified on each plot.}
    \label{fig:real-time reconfigurable}
  \end{center}
\end{figure}

The functionality of the reconfigurable electronics block is shown in Fig. \ref{fig:real-time reconfigurable}. A rectifier block generates the second harmonic used in the imaging process. The resulting signal is delayed in a delay line implemented in the DSP as a first-in-first-out queue. The amount of delay is proportional to the length of the queue, and is changed in real-time to implement the target functionality specified above. For the particular application described above, the phase delay for each metamaterial element indexed by variable $i=\overline{0,9}$ can be easily determined from the time needed by the acoustic energy to propagate between each element and the focal point at the central angular frequency, $\omega_0$, and it is given by

\begin{equation} \label{eq:lens equation}
 \phi_i(f, \alpha)=\phi_0-\dfrac{f\omega_0}{v}\sqrt{1+\left[\tan(\alpha)-\left(i-\frac{9}{2}\right)\frac{\Delta x}{f}\right]^2},
\end{equation}
where $f$ is the focal length, $\alpha$ is the angle of the focused beam, $\phi_0$ is a constant big enough to make all $\phi_i$ positive, $v=343$ m/s is the speed of sound, and $\Delta x=5$ cm is the metamaterial periodicity.

The supplemental movie \cite{Supplemental_Movie} shows the acoustic response of the slab to the three-pulse plane wave presented in Fig. \ref{fig:experimental setup}. Figure \ref{fig:real-time reconfigurable}b-d shows three representative frames separated by 13 ms that highlight the response of the lens for each  pulse. In each case, we note the clear convergent fields emerging from the lens, the focal spot marked on each  panel, and the focused beam leaving the scene in the designed direction, which confirm the desired functionality. The only deviation from the expected theoretical behavior is the side lobe visible in Figure \ref{fig:real-time reconfigurable}d and caused by the multiple reflections of the incident wave in the area between the lens and the speaker array.

\begin{figure}%[ht!]
  \begin{center}
    \includegraphics[width=8cm]{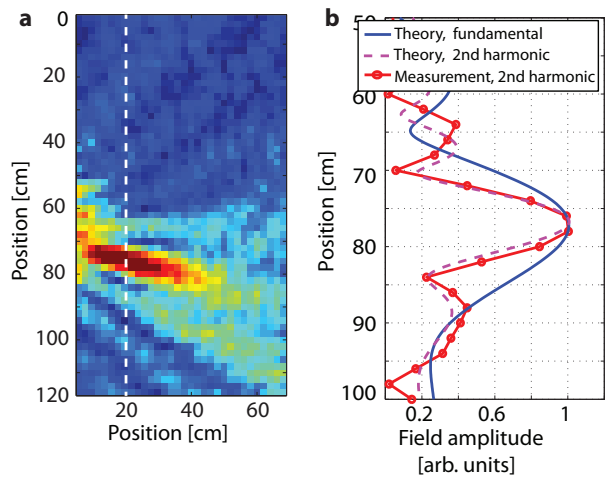}
    \caption{Second order harmonic lens performance. (a) Complex amplitude of the lens' second harmonic response (3000 Hz) to the third pulse shown in Fig. \ref{fig:real-time reconfigurable}d. The dotted line marks the focal plane. (b), Comparison between the field measured on the focal plane (circles) and an ideal lens imaging at the second harmonic (dotted curve) and the fundamental (solid curve). The second harmonic image is twice narrower than the image at the fundamental frequency.} 
    \label{fig:amplitude field}
  \end{center}
\end{figure}

These measurements emphasize the advantage of using higher order harmonics in imaging systems, namely the focal spot is considerably smaller than what can be obtained with regular, linear systems. To quantify the size of the focal spot, we present in Fig. \ref{fig:amplitude field}a the field amplitude at the second harmonic frequency of 3000 Hz measured for the lens' response to the second plane wave pulse. The field amplitude measured along the focal plane is plotted in Fig. \ref{fig:amplitude field}b as the circles marked curve and matches very well the theoretical prediction (dotted curve). Moreover, the image at the frequency of the incident field produced by an ideal linear lens having the same size and focal distance as the metamaterial lens, i.e. the diffraction limit, is shown using the continuous curve. We notice that the width of the measured image is approximately twice narrower than what is normally obtained in a linear system, in agreement with previous results reporting second harmonic imaging in the electromagnetic regime \cite{Wang_PRL_2011}.

In the above example, we showed how the active metamaterial can change functionality in real-time. Next, we show how it can be configured to serve multiple roles at the same time. To illustrate this idea, suppose we have a beam steering device that takes an incident beam and steers it to a different direction. Without disturbing this functionality, we want to monitor the quality of the beam by focusing it and measuring its characteristics using a microphone placed at the focal spot.

\begin{figure}%[ht!]
  \begin{center}
    \includegraphics[width=8.5cm]{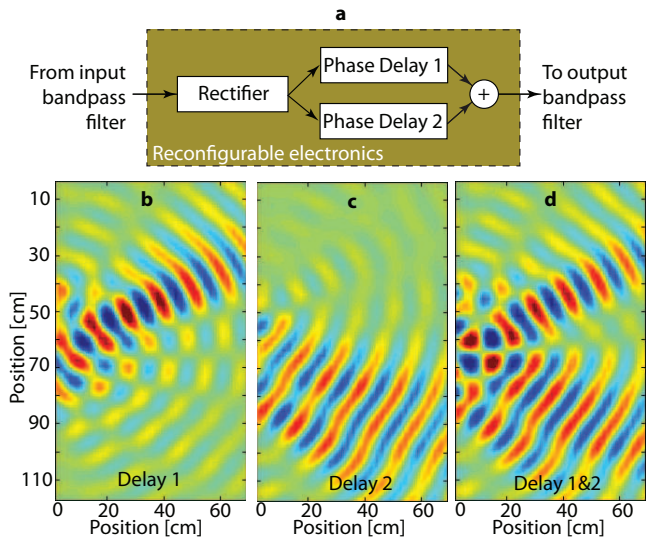}
    \caption{Metamaterial lens configured to multiplex functionality. (a) Diagram of the reconfigurable electronics showing two phase delay lines that implement the two desired behaviors. (b) The first functionality is that of a focusing lens identical to the one shown in Fig. \ref{fig:real-time reconfigurable}b. (c) Metamaterial configured to behave as a beam steering device. (d) Metamaterial configured to behave as a focusing lens and beam steering device simultaneously.}
    \label{fig:multiplexing functionality}
  \end{center}
\end{figure}

The double functionality is implemented in the reconfigurable electronics blocks as shown in Fig. \ref{fig:multiplexing functionality}a. The beam steering and focusing behavior is implemented using two delay lines implemented as before using first-in-first-out queues in the DSP. The signals leaving the queues are added and drive the metamaterial cell piezoelectric transducer that generates the cell acoustic response. 

The functionality implemented in the first delay line is that of a lens characterized by parameters (30 cm$, 30^\circ$), also used in the previous example. The beam steering functionality is obtained using Eq. (\ref{eq:lens equation}) and parameters $(f,\alpha)=(\infty,-30^\circ)$. To verify the correct behavior of each individual delay queue, we activated them individually and measured the acoustic response of the metamaterial slab to an incident plane wave. Figures \ref{fig:multiplexing functionality}a and \ref{fig:multiplexing functionality}b show the intended behavior of a lens and, respectively, beam steering device. By activating both delay lines we obtain the linear superposition of the two behaviors. More specifically, the plane wave and focused beams leaving the metamaterial slab are virtually identical to the beams of the basic devices whose functionality is combined, and confirm the excellent ability of the metamaterial slab to behave as two different devices at the same time.

We presented an active metamaterial design approach in which the effective material properties are set in reconfigurable digital electronic blocks and demonstrated the approach experimentally for a slab composed of ten unit cells. The design method features four main advantages over other acoustic metamaterial design methods. First, the material properties can be changed instantaneously to the desired values, which implies that metamaterials can be configured in real-time to have various behaviors depending on application without changing their physical structure. Thus, the slab was configured to individually manipulate three identical, closely spaced short pulses. Second, the metamaterial can be configured to play multiple roles at the same time. To this end, the metamaterial was configured to simultaneously behave as a beam steering device and focusing lens. Third, the design approach is suitable for non-linear imagining applications in which the device responds with higher order harmonics of the incident field. This is advantageous because it results in sharper images that do not obey the diffraction limit of linear systems. Finally, the metamaterial response is controlled almost entirely by electronics as opposed to the metamaterial geometry, which implies that it is now possible to obtain imaging devices much thinner than what is feasible with other methods. In one of the applications, the fabricated metamaterial slab was configured to behave as a lens a tenth of a wavelength thick and had an F-number below 0.5. These advantages demonstrate the versatility of the active metamaterial design procedure and recommend it to numerous applications.

\section*{Acknowledgments}
This work was supported by Grant No. N00014-12-1-0460 from the Office of Naval Research.

%\bibliography{../../acoustics.bib,../../metamaterials.bib,../../electromagnetics.bib}
%\bibliographystyle{plain}

\end{document}